# A NEW GENETIC CODE TABLE

Miloje M. Rakočević

*Department of Chemistry, Faculty of Science, University of Nis;*
*Home address: Milutina Milankovica 118, Belgrade: 11070, Serbia*
*(E-mail: m.m.r@eunet.yu ; www.sponce.net)*

**Abstract**

In this paper it is shown that within a Combined Genetic Code Table, realized through a combination of Watson-Crick Table and Codon Path Cube it exists, without an exception, a strict distinction between two classes of enzymes aminoacyl-tRNA synthetases, corresponding two classes of amino acids and belonging codons. By this, the distinction itself is followed by a strict balance of atom number within two subclasses of class I as well as two subclasses of class II of amino acids.

## 1 INTRODUCTION

Among the representations of the genetic code, two retain our attention here: the Watson-Crick Table, WCT (the standard genetic code Table) (Watson & Crick., 1953a,b; Crick, 1966, 1968), arranged through an ordering of codon letters that shows the 1st /2nd/ 3rd letters in the order UCAG/ UCAG/ UCAG (first permutation in Table 1), and the so-called Codon Path Cube, CPC (Swanson, 1984) with the order UCGA/UCGA/CUAG (second and third permutations in Table 1, respectively).

On the other hand, Wetzel (1995) has shown that a strict distinction of canonical amino acids (AAs) within the WCT in correspondence with two classes (I & II) of enzymes aminoacyl-tRNA synthetases (aaRS) is valid only for the X**U**X-X**C**X, but not for the X**A**X-X**G**X codon space[1]. Thus, as the AAs, handled by class I of aaRS within the X**U**X-X**C**X space, appear V, M, I, L whereas the AAs, handled by class II of aaRS are F, S, P, T, A. [Eriani et al., 1995, p 499: "The class I enzymes chare with dehydrogenases and kinases *the classic* nucleotide binding *fold* called the Rossmann fold, whereas the class II enzymes possess *a different fold, not found elsewhere*, built around a six-stranded antiparallel beta-sheet. The two classes of synthetases ... differ as to where on the terminal adenosine of the tRNA the amino acid is placed: class I enzymes act on the 2′ hydroxyl whereas the class II enzymes prefer the 3′ hydroxyl group”. (My *italics* and my comment: the C atom on the position 2′ is a stronger nucleophile whereas the C atom on the position 3′ is a less strong nucleophile).]

[1] Wetzel, 1995, p. 545: "all XCX codons code for amino acids handled by class II synthetases, and all but one of the XUX codons code for amino acids handled by class I synthetases"; cf. also Fig. 1 in the cited work on the page 546.

| | | |
|---|---|---|
| **1** | **UCAG** / GACU | **4** |
| **2** | UCGA / AGCU | **3** |
| **3** | CUAG / GAUC | **2** |
| **4** | CUGA / AGUC | **1** |
| | | |
| **1** | **UACG** / GCAU | **4** |
| **2** | UAGC / CGAU | **3** |
| **3** | AUCG / GCUA | **2** |
| **4** | AUGC / CGUA | **1** |
| | | |
| **1** | UGCA / ACGU | **4** |
| **2** | UGAC / CAGU | **3** |
| **3** | GUCA / ACUG | **2** |
| **4** | GUAC / **CAUG** | **1** |

**Table 1** The 12+12 permutations of the sequence of four amino bases, two of pyrimidine (Py) and two of purine (Pu) type, arranged through three octets, in accordance to three main characteristics (footnote 4) of amino bases. Only first three permutations are included in CT.

It is also shown (Rakočević, 1997a,b) that within the CPC (Figure 1 in Rakočević, 1997a) the mentioned distinction is valid for the whole X**U**X-X**C**X-X**A**X-X**G**X codon space, *with* only one exception. Namely, within space I there are AAs: M, I, V, L, Y, Q, E, C, W and pyrimidine-coding arginine R, all handled by class I of aaRS, and within space II: F, S, P, A, T, G, K, N, D, H, all handled by class II, and purine-coding arginine R, handled by class I of aaRS.

In this paper, however, we show that within a Combined Genetic Code Table (CT), combined in a specific manner from the WCT and the CPC (Table 2 in relation to Table 3), the strict distribution of the AAs over the two classes of aaRS occurs *without* an exception.

| 2nd letter | 1st letter: *A* / *G* | | *G* / *A* | | *C* / *C* | | *U* / *U* | | 3rd letter |
|---|---|---|---|---|---|---|---|---|---|
| *A* | **AAC** | **N** | **GAC** | **D** | **CAC** | **H** | UAC | | *C* |
| | **AAU** | | **GAU** | | **CAU** | | UAU | Y | *U* |
| | **AAA** | | GAA | | CAA | | UAA | | *A* |
| | **AAG** | **K** | GAG | E | CAG | Q | UAG | * | *G* |
| *G* | **GGA** | | AGA | | CGA | | UGA | * | *A* |
| | **GGG** | | AGG | R | CGG | | UGG | | *G* |
| | | **G** | | | | R | | W | |
| | **GGC** | | **AGC** | | CGC | | UGC | | *C* |
| | **GGU** | | **AGU** | **S** | CGU | | UGU | C | *U* |
| *C* | **GCA** | | **ACA** | | **CCA** | | **UCA** | | *A* |
| | **GCG** | **A** | **ACG** | **T** | **CCG** | **P** | **UCG** | **S** | *G* |
| | **GCC** | | **ACC** | | **CCC** | | **UCC** | | *C* |
| | **GCU** | | **ACU** | | **CCU** | | **UCU** | | *U* |
| *U* | AUC | | GUC | | CUC | | **UUC** | | *C* |
| | AUU | I | GUU | V | CUU | L | **UUU** | **F** | *U* |
| | AUA | | GUA | | CUA | | UUA | | *A* |
| | AUG | M | GUG | | CUG | | UUG | L | *G* |
| 2nd letter | *G* / *A* | | *A* / *G* | | *C* / *C* | | *U* / *U* | | 3rd letter |
| | 1st letter | | | | | | | | |

**Table 2.** A new Table of genetic Code as a Combined Table, realized in a specific combination of Watson-Crick Table and Codon Path Cube (Swanson, 1984), through the first three permutations from the first octet in Table 1. Bold positions (dark tones): the codons coding for AAs handled by class II of aaRS; non-bold positions: the codons coding for AAs handled by class I aaRS plus three "stop" codons, denoted with *.

| 2nd letter | 1st letter | | | | | | | | 3rd letter |
|---|---|---|---|---|---|---|---|---|---|
| | *A* *G* | | *G* *A* | | *C* *C* | | *U* *U* | | |
| *A* | **AAC** | **N** | **GAC** | **D** | **CAC** | **H** | UAC | | *C* |
| | **AAU** | | **GAU** | | **CAU** | | UAU | Y | *U* |
| | **AAG** | | GAG | | CAG | | UAG | | *G* |
| | **AAA** | **K** | GAA | E | CAA | Q | UAA | * | *A* |
| *G* | **GGA** | | AGA | | CGA | | UGA | * | *A* |
| | **GGG** | | AGG | R | CGG | | UGG | | *G* |
| | | **G** | | | | R | | W | |
| | **GGU** | | **AGU** | | CGU | | UGU | | *U* |
| | **GGC** | | **AGC** | **S** | CGC | | UGC | C | *C* |
| *U* | AUG | M | GUG | | CUG | | UUG | | ***<u>G</u>*** |
| | AUA | | GUA | V | CUA | L | UUA | L | ***<u>A</u>*** |
| | AUU | I | GUU | | CUU | | **UUU** | | ***<u>U</u>*** |
| | AUC | | GUC | | CUC | | **UUC** | **F** | ***<u>C</u>*** |
| *C* | **GCC** | | **ACC** | | **CCC** | | **UCC** | | *C* |
| | **GCU** | **A** | **ACU** | **T** | **CCU** | **P** | **UCU** | **S** | *U* |
| | **GCG** | | **ACG** | | **CCG** | | **UCG** | | *G* |
| | **GCA** | | **ACA** | | **CCA** | | **UCA** | | *A* |
| 2nd letter | *A* *G* | | *G* *A* | | *C* *C* | | *U* *U* | | 3rd letter |
| | 1st letter | | | | | | | | |

**Table 3.** A variant of Combined genetic code table, realized through *all four* left permutations from first octet in Table 1. Bold positions (dark tones): the codons coding for AAs handled by class II of aaRS; non-bold positions: the codons coding for AAs handled by class I of aaRS. Notice that fourth permutation (CUGA), in determination of third codon position, can be used only in three rows because the "middle U" is a "gooseneck" – the distinction within three isoleucine codons is not allowed. (About relationships between Table 2 & 3, *see* in Discussion.)

The first subclass of class I (1 in Table 4) is very down in Table 2 and the second one (1’ in Table 4) in the top, on the right. On the other hand, the first subclass of class II (2 in Table 4) is in a middle space (below in Table 2) whereas the second one is above on the left (2’ in Table 4).

| | | | | | | | | | | | | | | | | | |
|---|---|---|---|---|---|---|---|---|---|---|---|---|---|---|---|---|---|
| 1 | 11 | + | 39 | + | 40 | + | 78 | = | 168 | | | | | | | | |
| | **M** | + | **I** | + | **V** | + | **L** | = | **(4)** | | | | | | | | |
| 2 | 32 | + | 16 | + | 32 | + | 20 | + | 28 | = | 128 | | | | | | |
| | **T** | + | **A** | + | **P** | + | **S** | + | **F** | = | **(5)** | | | | | | |
| 2’ | 16 | + | 14 | + | 22 | + | 30 | + | 10 | + | 04 | = | 096 | | | | |
| | **N** | + | **D** | + | **H** | + | **K** | + | **S** | + | **G** | = | **(6)** | | | | |
| 1’ | 30 | + | 22 | + | 20 | + | 34 | + | 68 | + | 10 | + | 18 | = | 202 | | |
| | **Y** | + | **Q** | + | **E** | + | **R** | + | **R** | + | **C** | + | **W** | = | **(7)** | | |

**Table 4**. The atom and molecule number balance. The two (1 & 1’) and two (2 & 2’) rows correspond to two and two subspaces (subclasses) within the Combined Table, presented here in Table 2.

**Survey 1.** *The logic of codon position determination (with I, II and III permutations) in Table 2**

| **Perm.** | **Posit.** | **Rows** |
|---|---|---|
| II | 2 | 2 outer |
| | 2 | 2 inner |
| | 1 | 2 outer |
| | 3 | 2 inner |
| III | 3 | 2 outer |
| I | 1 | 2 inner |

*The order: outer/inner/outer/inner

**Survey 2.** *Atom number balances*

| Nucleotides | | Amino acids (Tab. 4) |
|---|---|---|
| **U**MP (34) + **G**MP (38) = 72 ± 0 | | |
| **C**MP (35) + **A**MP (37) = 72 ± 0 | | |
| | 73 – 1 = | 168 – 96 (1 – 2') |
| **U**MP (34) + **A**MP (37) = 72 – 1 | | |
| **C**MP (35) + **G**MP (38)= 72 +1 | | |
| | 73 + 1 = | 202 – 128 (1' – 2) |

As we can see from Table 4, the calculation of the number of atoms within amino acid molecules (side chains), requires that each molecule be included as many times as there are codons coding for it. By this one must notice the situation R + R (meaning that R occurs two times) in row 1' because arginine possesses two spaces within the WCT; and the situation with only one L in row 1 because Leucine possesses only one space (through six neighbor codons) within the same Table. Certainly, serine also must appear two times (S in space "C" in row 2 and S in space "G" in row 2') in accordance with its two spaces within the WCT[2].

In addition, this CT shows the existence of a strict balance of the number of atoms within the amino acid molecules (their side chains) according to the *principle of minimum change* (Table 4 in relation to Surveys 2 & 3). As it will be shown below, the word is about the changes exactly for ±0 or ±1 of atoms in relation to arithmetic mean for two and two amino acid subclasses.

Together with all these balances, there is a balance of the number of molecules: 1+1' = 2+2', i.e. 4+7 = 5+6 (where the principle of minimum change is also valid through the sequence of the 4-5-6-7).

[2] In contrary, in Shcherbak's nucleon number balance between four-codon and non-four-codon amino acids (Shcherbak, 1994; 2003) Leucine possesses two positions too, because it is distributed into two four-codon families. This fact affirms a conclusion that multi-meaning states are from the inherent characteristics for the genetic code.

**Survey 3.** *Relations within Table* 4

(1 + 1') 168 + 202 = **297** + <u>073</u> = <u>370</u>

(2 + 2') 128 + 096 = **297 -** <u>073</u> = 224

(1 + 2') 168 + 096 = 8 x 33

(1' + 2) 202 + 128 = 10 x 33

(1 + 2) 168 + 128 = (9 x 33) **-** 1

(1' + 2') 202 + 096 = (9 x 33) + 1

9 x 33 = **297**

| | |
|---|---|
| [(Py-Py)U]: $F_2$ (28) + $L_2$ (26) = 54 | [(Py-Pu)U]: $L_2$ (26) + $L_2$ (26) = 52 |
| [(Py-Py)C]: $S_2$ (10) + $P_2$ (16) = 26 | [(Py-Pu)C]: $S_2$ (10) + $P_2$ (16) = 26 |
| [(Py-Py)A]: $Y_2$ (30) +$H_2$ (22) = 52 | [(Py-Pu)A]: ** (00) + $Q_2$ (22) = 22 |
| [(Py-Py)G]: $C_2$ (10) + $R_2$ (34) = 44 | [(Py-Pu)G]:*W(18) + $R_2$ (34) = 52 |
| [(Pu-Py)U]: $I_2$ (26) + $V_2$ (20) = 46 | [(Pu-Pu)U]: IM (24) + $V_2$ (20) = 44 |
| [(Pu-Py)C]: $T_2$ (16) + $A_2$ (08) = 24 | [(Pu-Pu)C]: $T_2$ (16) + $A_2$ (08) = 24 |
| [(Pu-Py)A]: $N_2$ (16) +$D_2$ (14) = 30 | [(Pu-Pu)A]: $K_2$ (30) + $E_2$ (20) = 50 |
| [(Pu-Py)G]: $S_2$ (10) + $G_2$ (02) = 12 | [(Pu-Pu)G]: $R_2$ (34) + $G_2$ (02) = 36 |

**Table 5.** The distribution of the canonical AAs in correspondence to the first-third-letter codon rule (Siemion & Siemion, 1994) in WCT, over the number of atoms within amino acid molecules (side chains). The index designates the number of codons, coding for the given amino acid.

| I - III | U | C | A | G | | | |
|---|---|---|---|---|---|---|---|
| Py – Py | 54 | 26 | 52 | 44 | 176 | | 330 |
| Py – Pu | 52 | 26 | 22 | 52 | 152 | 330-066 | |
| Pu – Py | 46 | 24 | 30 | 12 | 112 | | |
| Pu – Pu | 44 | 24 | 50 | 36 | 154 | | |
| | 196 | 100 | 154 | 144 | | | |
| | **[(330-033) –1]** | | **[(330-033) +1]** | | | | |

**Table 6**. The atom number balances within WCT after first-third-letter codon position rule (Siemion & Siemion, 1994) expressed through the number of atoms within amino acid molecules (side chains). First column (I-III) designates the type of the base in first-third position of corresponding codons. The letters U, C, A, G are related to four columns in WCT. Within two inner and two outer rows as well as within two first and two second columns there are (8 x 33), [(9 x 33)±1], and (10 x 33) of atoms, respectively.

From Survey 2 it follows that *the differences* in the number of the atoms within amino acid subclasses 1' and 2 (202-128) equals 73+1 and within subclasses 1 and 2' (168-96) equals 73 –1 or, in other words, both in relation to arithmetic mean 73±1. On the other hand, *the sums* of the atoms within two and two types of nucleotide molecules equal as follows; in first case: **C**MP + **G**MP = 72+1 and **U**MP + **A**MP = 72 –1, or, both in relation to arithmetic mean 72±1; and in second case: **C**MP + **A**MP = 72±0 and **U**MP + **G**MP = 72±0, or, both in relation to arithmetic mean 72±0.

In Survey 3 are presented some new balances from which it follows that the distribution into 8 x 33; 9 x 33 (±1) and 10 x 33 of atoms appears to be the same as in Py/Pu distinction within WCT from the aspect of "first-third-letter codon position rule" (Siemion & Siemion, 1994)[3] (see Tables 5 and 6 in this paper in relation to Tables 3a and 3b in Rakočević, 2004).

[3] Siemion and Siemion, 1994, p. 139: "It is shown that in the pairs of amino acids coded by the codon possessing identical bases in the first and second positions, the amino acid with R in the third position are of higher structural importance than the amino acids coded with Y" (*Remark* 1:

## 3 DISCUSSION

The key for understanding the arrangement in this new Genetic Code Table, as it is given in Table 2, is the alphabet-permutation hierarchy, given in Table 1.The first 4+4 permutations are given in relation to the type of the base (to be Py or Pu), next 4+4 in relation to the number of hydrogen bonds (realized between codon – anticodon), and the last 4+4 in relation to the functional group (to be *oxo*, i.e. hydroxyl, or *amino* functional group in terminal molecule position)[4]. By this, one must notice that permutation arrangement is given through the logic of existence *the pairs of the pairs*, and also notice that it makes sense to determine the permutation ordinal number only within each three octets, in two manner: going from up to down, on the left, and in a vice versa direction, on the right. (*Remark* 2: It is necessary to distinguish the principle "*to be the pair of the pairs*" from the principle "*to exchange a pair for a pair*"; cf. Remark 3.)

That follows from the mirror symmetry, which can be realized through two logics: 1. Single letter to single letter, and 2. Letter doublet to letter doublet. Thus, as the first permutation in third octet, we have: in first logic UGCA / ACGU and in second one: UGCA / CAUG. That is the reason why the permutation CAUG is the first, and not – the fourth[5].

In the other words, the key problem in this analysis of the relationships in CT is the question which is related to the status of permutations within Table 1. Bearing in mind this aspect of the said question, and in order for better understanding, we mention here some additional facts. So, the second permutation in first octet (UCGA) is also used as a codon arrangement in WCT, when occurs a strict distinction between four-codon and non-four-codon AAs (Yang, 2003).

---

Except with Pu, a purine can be denoted with the letter R, and a pyrimidine, Py, with the letter Y as it is here.)

[4] Rakočević, 1988, p. 112: „A system of four „small" molecules is made up of two purine and two pyrimidine bases. ... The four [molecules] are mutually distinguishable by three main characteristics: the type of base (purine, Pu, or Pyrimidine, Py); the type of functional group in the terminal position (position 6 in purine, position 4 in pyrimidine) – either oxo or amino; and the number of hydrogen bonds linking them in the system codon – anticodon."

[5] This cyclization itself is also valid for all "8 x 3" quadruplets in three permutation octets. The quadruplet UUXX is realized three times going in up/down, and three times in a vice versa direction; so on the left, and so on the right, regarding the whole permutation system (two and two outer columns). The same is valid for the quadruplets XYXY.

On the other hand, two permutations from third octet appear also as very characteristic, and in a specific connection through Py-pair / Pu-pair inversion, both on the right and in ordering read from down to up; the first and fourth permutations, CA-UG and AC-GU, respectively. Namely, the CA-UG permutation is used in a p-adic mathematics application to the genetic code structure (Dragovich & Dragovich, 2006), whereas from the AC-GU it follows an ordinal number of AAs within WCT from 0 to 19 (Damjanović, 1998; Damjanović & Rakočević, 2005, 2006).

From this follows that two permutations, CA-UG and AC-GU appear as a new pair (through an inversion), the outer pair, complementary with the inner pair CA-GU and AC-UG. The same is valid for the first quadruplet (and for all other), and that is the reason why the fourth permutation (CUGA) is so power for a new splitting into the CT, such a splitting from that follows Table 3 as "a freedom degree" for CT, presented in Table 2. (*Remark* 3: The existence of one outer and one inner pair within eight quadruplets, in Table 1, one can understand as the principle – *to exchange a pair for a pair*; cf. Remark 2.)

In accordance to this discussion there are some possible chemical reasons, from which we give here only a choice. Namely, we will give only an explanation for the ordinal number 1, going from the simplest to the most complex of amino acid molecules; complex from different aspects (three toned spaces in Table 1).

*

*First octet, permutation UCAG*: From the aspect of the type of the base, pyrimidine is simpler than purine (*one* ring versus *two* rings). On the other hand, U and A with *two* hydrogen bonds both, are simpler than C and G, which possess *three* hydrogen bonds each.

*Second octet, permutation UACG*: From the aspect of the number of hydrogen bonds, U and A are simpler than C and G. On the other hand U is simpler than A as a pyrimidine molecule.

*Third octet, permutation CAUG*: From two purines A is simpler than G, because it possesses an amino functional group only; in the other words, G is more complex through its two functional groups – amino and oxo[6]. (In addition, one must notice that nitrogen is simpler than oxygen.) In a pairing process, the base U, possessing two more complex functional groups (the oxo group, two times!) must go with G, and simpler C, with only one oxo group, must be with A. Certainly, in both cases of pairing, the pyrimidine must be before purine as the simpler molecule.

[6] The base A, as we know, is simpler than G through the number of hydrogen bonds too.

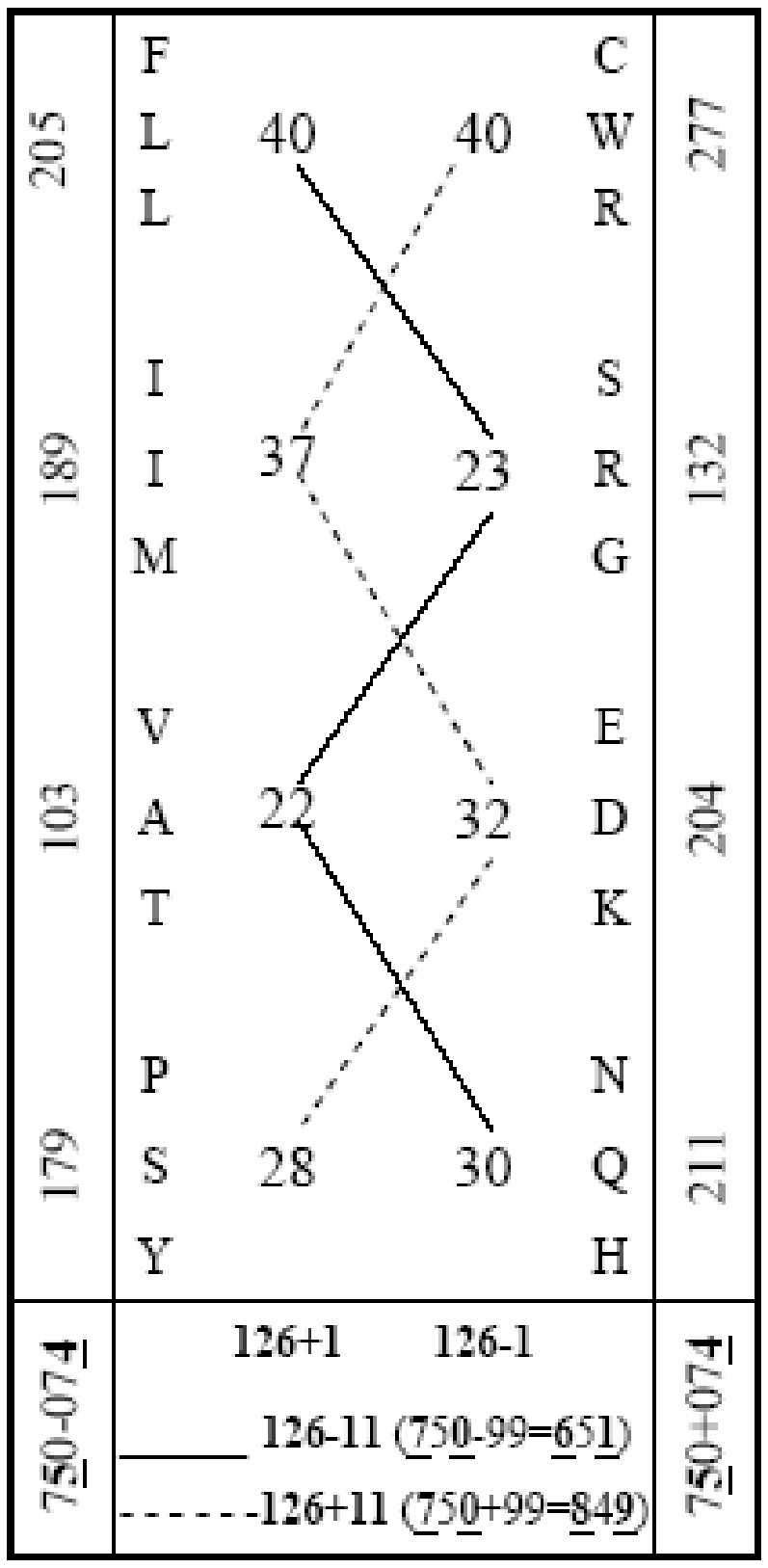

**Table 7.** The distribution of the canonical AAs in standard Genetic Code Table, in correspondence with the Py-Pu distinction, atom and nucleon number (unit change) balance (including the balance of the isotope number as it is shown in Remark 5)[7]. Notice that AAs are given in ordering as they exist within standard GCT, going from column to column. Notice that amino acid arrangement exists as 12±1 AAs between two "stop" codon positions. First 12 AAs before UAA (ochre) and UAG (amber); and next 12-1 before UGA (opal). Reading from C over F there are 12+1 AAs. Notice that quantum "074" is a head-nucleon-number.

In connection with the first, there is a second question which is related to the hierarchy of the three permutation octets. After our hypothesis, the hierarchy must be – as it is given. That is so, because such a hierarchy is found in Gray code model of genetic code (Swanson, 1984) as well as on the genetic code binary tree (Rakočević, 1998)[8].

[7] The reading logic, e.g. for Proline: CCY and CCR – one and the same amino acid, P; e.g. for Isoleucine: AUY one amino acid – I; AUR – two different AAs: I and M; in total: I, I, M. (The letters Y & R as said in Remark 1.)

*

*Remark* 4: Atom number and nucleon number within amino acid side chains: F 14 (91), L 13 (57), I 13 (57), M 11 (75), V 10 (43), A 04 (15), T 08 (45), P 08 (41), S 05 (15), Y 15 (107), H 11 (81), Q 11 (72), N 08 (58), K 15 (72), D 07 (59), E 10 (73), G 01 (01), R 17 (100), W 18 (130), C 05 (47).

*Remark* 5: The isotope (nuclide) number balance in Table 7: the difference between left and right half equals **±0**, as follows: F28 + L26 + L26 + I26 + I26 + M24 + V20 + A08 + T17 + P16 + S11 + Y31 = 259; 259 = H22 + Q23 + N17 + K30 + D16 + E22 + G02 + R34 + S11 + R34 + W36 + C12.

*Remark* 6: The isotope (nuclide) number balance in Table 8, in a zigzag (periodic) ordering: the difference between two halves equals **±1**, as follows: (F28 + L26 + S11) + (A08 + V20 + R34) + (L26 + P16 + H22) + (T17 + M24 + I26) = 259–1; 259 + 1 = (G02 + E22 + D16) +(Y31 + C12 + W36) + (S11 + K30 + N17) + (Q23 + R34 + I26).

*Remark* 7: The number 259 is the first permutation in column 7 of Shcherbak's modular system (in module 9) of multiples of number 037 (Table 1 in Shcherbak, 1994, p. 476); moreover, the nucleon number within four-codon and non-four-codon AAs is determined only with 7$^{th}$ column, as multiples of 259, and with 1$^{st}$ row, as multiples of 111.

*Remark* 8: An example of calculation of particles number for serine side chain: atom number: $CH_2$ + OH = 5 atoms; isotope number: C-12 and C-13 equals 2 isotopes; H-1 and H-2 (three times) equals 6 isotope; O-16, O-17 and O-18 equals 3 isotopes; in total: 2 + 6 + 3 = 11 isotopes; nucleon number within first isotope, i.e. nuclide: 12 + (3 x 1) + 16 = 31.

---

[8] On the Gray code model as well as on the genetic code binary tree, the first codon classification is given after type of the base (Py/Pu distinction) and second one after the number of hydrogen bonds. The logic existing within third characteristics (functional group *oxo* or *amino* in terminal Py/Pu base position) parallel with these two classifications is not found. From this it follows: neither to be the first, nor the second one, then must be – the third (the third octet of permutations in Table 1).

| | | | | | |
|---|---|---|---|---|---|
| 179 | F<br>L<br>S | 32 | 18 | G<br>E<br>D | 133 |
| 284 | Y<br>C<br>W | 38 | 31 | A<br>V<br>R | 158 |
| 179 | L<br>P<br>H | 32 | 28 | S<br>K<br>N | 161 |
| 229 | Q<br>R<br>I | 41 | 32 | T<br>M<br>I | 177 |
| 750+121<br>871 | 126+017 126-017<br>______ 126-1 (693)<br>------126+1 (807) | | | | 750-121<br>629 |

**Table 8.** The distribution of the canonical AAs in standard Genetic Code Table, in correspondence with the Py-Pu distinction, atom and nucleon number (unit change) balance, including the balance of the isotope number as it is shown in Remark 6. Notice that AAs are given in ordering as they exist within standard GCT, going from row to row. Notice $121=11^2$.

In some manner, independently from discussed questions, there is a fundamental question, which is related to the "choice" of the number of letters within a code-alphabet and, *per se*, the number of permutations. It is self-evident that from the aspect of the principle "to be pair of pair", the 3-letter, 5-letter, 6-letter and 7-letter alphabets are not possible. The next, after the 4-letter alphabet, which seems to be possible, is just the 8-letter alphabet. But that is an illusion. Why? In 4-letter alphabet each permutation really contains a pair of pair of letters: (a-b, c-d) whereas in 8-letter alphabet – more than a pair of pair. From this aspect, and from the aspect of the above discussed two principles "the pair of the pairs" and "a pair for a pair" (Remarks 2 & 3), it makes sense to conclude that the system of 24 permutations (12 pairs), given in a manner as in Table 1, is one and only possible solution. Bearing this state in mind, it is clear why the amino acid corresponding Py-Pu codon splitting, in standard Genetic Code Table, is also given in 24 amino acid meanings (Tables 7 and 8), i.e. as 12 pairs in both cases[9].

## CONCLUSION

In conclusion we express a hope that the revelation of these regularities can help in a better understanding of codon - amino acid assignment over two classes of aaRS (De Duve, 1988; Hipps, 1995; Schimmel, 1995). On the other hand, such strict connections between two classes of enzymes, two classes of AAs and belonging codons, provide evidence to support the hypothesis, given in a previous paper (Rakočević, 2004), that genetic code was complete from the very beginning as the condition for origin and evolution of the life.

---

[9] Rakočević & Jokić, 1996, p. 346: "Notice that out of all doublet-triplets systems, this is the only and one with two possible distinctions for doublets (to be six and six, and then, to be three and three doublets) and three possible distinctions for triplets (to be four and four, then two and two, and, finally, to be one and one triplet)".